\documentclass[aps,prd,twocolumn,amsmath,amssymb,showpacs,showkeys]{revtex4}
\usepackage{graphicx}
\usepackage{subfigure}
\usepackage{bm}

\newcommand{\be}{\begin{equation}}
\newcommand{\ee}{\end{equation}}
\newcommand{\bea}{\begin{eqnarray}}
\newcommand{\eea}{\end{eqnarray}}
\def\eq#1{(\ref{#1})}

\newcommand{\nn}{\nonumber\\}
\def\mr#1{{\mathrm{#1}}}

\begin{document}
\title{Exchange of signals around the event horizon in Schwarzschild space-time}
\author{M. Gawe{\l}czyk$^a$, J. Polonyi$^b$, A. Radosz$^a$, A. Siwek$^{ab}$}
\affiliation{$^a$ Institute of Physics, Wroc{\l}aw University of Technology, Wybrze\.ze Wyspia\'nskiego 27, 50-370 Wroc{\l}aw, Poland}
\affiliation{$^b$University of Strasbourg,
High Energy Physics Theory Group, CNRS-IPHC,
23 rue du Loess, BP28 67037 Strasbourg Cedex 2, France}

\begin{abstract}
Red shift in communication and possibility of interaction is discussed for objects around
the event horizon of Schwarzschild space-time. It is pointed out that the arrow of time
within the horizon cannot always be inferred by observations carried out outside. Two scenarios
are presented for the causal structure of the space-time and it is found that in one of 
them extended objects fall apart into their elementary constituents by crossing the horizon.
\end{abstract}

\date{\today}

\maketitle

\section{Introduction}
The space-time coordinates are extracted from particles' world-lines by inspecting their motion. The special dynamical feature of time in classical physics is its 
orientation, the arrow of time \cite{Zeh}. The coordinate of a particle may increase and 
decrease as a function of the proper time of the world-line but the time is a monotonic
function in classical physics. It will be pointed out in this paper that the choice of the
arrow of time and the choice of the time coordinate is nontrivial within the event horizon 
of Schwarzschild space-time.

The event horizon in itself is the source of a number of interesting phenomena, thermodynamical properties
\cite{Bardeen,Bekenstein} and particle creation \cite{Hawking} being the most interesting and debated 
questions (see also \cite{Leonhardt}). Interesting features may be revealed in the context of 
communication or in general interactions in the vicinity of the event horizon. Below we 
discuss issues related to the necessity of choosing an arrow of time. To our best 
knowledge the possible change in causal structure, a significant issue of the 
present work, has not been discussed before.

It is well known that tidal forces (and other physical phenomena) are regular at the event horizon of the Schwarzschild geometry \cite{Lematire}. Nonetheless, the horizon is a special surface, for instance it is a separatrix for geodesics which emanate from the center. It is pointed out that the causal structure, namely the direction of the arrow of time may or may not be regular at the horizon on the outgoing massless geodesics. Note that such an ambiguity causes no physical singularity itself, but still two possible below-horizon scenarios arise. Another specific feature of the horizon which may be important in determining the causal structure is that this surface, $r=r_S$, which seems to be three-dimensional, has zero volume, as $g_{tt}=0$ (see e.g. discussion in \cite{Misner}).

Communication by means of electromagnetic signals exchanged between observers traveling 
through a gravitational field has already been considered in various other aspects. Discussion in this area has been focused on two opposite regimes. The weak-field, 
slow-motion regime relates to its applications in the frame of Satellite Navigation and 
Communication Systems \cite{Coll,Coll2,Linet,Bahder}. In the regime of strong gravitational fields, 
varieties of the consequences of the presence of the event horizon have been investigated. Namely, 
problems of speed at the crossing instant as reaching the value of the speed of light 
\cite{Janis}, information carried away by an object crossing the horizon \cite{Dragan}, the 
history of the universe as seen by a falling observer crossing the horizon \cite{Grib} and some geometrical aspects of distortion of the  stellar sky by a black hole \cite{Muller,Muller2} have been considered. An interesting aspect of communication between accelerated observers was recently discussed in Ref. \cite{Friis}. It was shown that infinite acceleration (effectively) degrades quantum entanglement both in bosonic and fermionic cases. In this way, though not directly as in the case of uniform acceleration discussed in \cite{Friis}, the issue of the role of a strong gravitational field for quantum information theory is raised. The role of the Schwarzschild event horizon in the process of matching propagators in the exterior and interior of a black hole, resulting in particle production has recently been discussed in Ref. \cite{Tsoupros}. The papers \cite{Friis} and \cite{Tsoupros} belong to different sectors of relativistic context of quantum information and communication, while the rest of this paper is devoted to the non-quantum aspect of communication near the event horizon.

Our aim in this paper is to enlighten the problem of the exchange of electromagnetic signals in the case when one or both observers are crossing the horizon. In this latter case, one can (naively) expect disruption of communication when one of the observers crosses the horizon. One finds instead an anomaly in the perception of this fact by the other observer, the chasing one. Such an anomaly is accompanied by an irregularity in the ``outgoing'' signals: outgoing radial null geodesics appear to be ill defined on the horizon. In consequence, the extension of this type of geodesics below the horizon may not be a unique one and two possible alternatives for the communication in the black hole interior arise. We will 
briefly characterize two scenarios corresponding to two different orientations of the time flow in this case.
An important context of our discussion is to point out that below the horizon the arrow of time on the outgoing geodesic cannot be determined by observations carried out in the black hole exterior.

This discussion is made by tracing the exchange of light signals in the vicinity of 
the event horizon among three observers: observer A (Alice), observer B (Bob), and
observer ms (mother station). For simplicity we shall assume that both Alice and Bob, 
the latter following the former, originate from the same ms, arranged at radial position 
$r=r_{ms}$ in Schwarzschild coordinates. Two particular questions will be discussed. How does the 
event horizon influence ms - Alice's communication, as recorded by Alice, in the black hole 
interior and how does it affect the two-way communication between Alice and Bob, which is expected 
to be broken when Alice crosses the event horizon. The analysis both in exterior and interior 
of the black hole is performed within Kruskal-Szekeres coordinates \cite{Kruskal,Szekeres}. One finds 
that Alice receives signals from  ms with a universal redshift value of $1/2$ at the event 
horizon. Two-way A-B communication turns out to be disrupted in a special, rather 
unexpected way being inevitably broken and never restored in one scenario and being 
restored below the horizon in the other scenario. A simple measure of the loss of
two-way communication is introduced and monitored during a test object falling
through the horizon (in the first scenario).

The Kruskal-Szekeres coordinates are introduced in Section II, followed by a
short presentation of the redshift observed by communicating above the
horizon. It is pointed out in Section III that the time arrow of geodesics shoving
away the center within the black hole cannot be determined by observations carried
out outside. As a result, we cannot distinguish between two scenarios, differing
in the signature of the time coordinate within the black hole. A measure of 
the lack of two-way communication in one of these scenarios is presented in Section IV.
Finally, Section V is devoted to our conclusions.

\section{Above the horizon}
Communication above the horizon in principle may be described by means of the Schwarzschild coordinate
system equipped with the metric
\be\label{schw}
ds^2=\left(1-\frac{r_S}{r}\right)dt^2-\frac1{1-\frac{r_S}{r}}dr^2-r^2d\Omega^2,
\ee
where $r_S=2M$ ($c=G=1$) denotes critical radius (event horizon). However, in order to 
preserve coherent approach for interior and exterior regions, here we apply Kruskal-Szekeres
coordinates \cite{Kruskal} defined by the transformation $(t,r)\to(u,v)$:
\bea\label{schkr}
u&=&\sqrt{\left|\frac{r}{r_S}-1\right|}e^{\frac{r}{4M}}\begin{cases}\cosh\frac{t}{4M}&r>r_S\cr
\sinh\frac{t}{4M}&r<r_S\end{cases},\nn
v&=&\sqrt{\left|\frac{r}{r_S}-1\right|}e^{\frac{r}{4M}}\begin{cases}\sinh\frac{t}{4M}&r>r_S\cr
\cosh\frac{t}{4M}&r<r_S\end{cases}
\eea
for $u+v>0$ with a line element
\be
ds^2=K(dv^2-du^2)-r^2d\Omega^2,
\ee
where $K=32M^3\exp(-r/2M)/r$. The inverse transformations are given by
\bea\label{inverse}
\left(\frac{r}{r_S}-1\right)e^{\frac{r}{2M}}&=&u^2-v^2,\nn
\mr{atanh}\frac{t}{4M}&=&\begin{cases}\frac{v}{u}&r>r_S\cr\frac{u}{v}&r<r_S\end{cases}.
\eea
Exterior of the black hole corresponds to $u>v$. Then the velocity vector $V^\mu V_\mu=1$ for massive (radial) geodesics has the following $u,v$ components:
\bea\label{velkr}
V^v&=&\frac{4M}{K(u^2-v^2)}\left[uE-v\sqrt{E^2-\frac{K}{16M^2}(u^2-v^2)}\right],\nn
V^u&=&\frac{4M}{K(u^2-v^2)}\left[vE-u\sqrt{E^2-\frac{K}{16M^2}(u^2-v^2)}\right],
\eea
where the parameter $E$ is determined by the function
\be\label{hfnct}
h(u,v)=\frac{\sqrt{K(u^2-v^2)}}{4M}
\ee
at the initial conditions, $E=h(u_{ms},v_{ms})$.

Putting E=1 in Eq. \eq{velkr}, which corresponds to fall from infinity, one reproduces the solution of Ref. \cite{Trumper}. 
The vector tangent to massless geodesic, $k^\mu k_\mu=0$, has the components

\be\label{in}
k_+^v=- k_+^u=\frac{4M\Omega}{K(u+v)}.
\ee
\be\label{out}
k_-^v= k_-^u=\frac{4M\Omega}{K(u-v)}.
\ee

The indices + and - stand for the ingoing and outgoing signals, respectively; 
these null geodesics are straight lines inclined by $\mp\pi/4$ to the coordinate axis $u$.
The parameter $\Omega$ represents the frequency as measured by a hypothetical observer in rest
at spatial infinity if the light signal goes to or comes from infinity. 

The particular feature of Schwarzschild geometry is the presence of the 
absolute rest frame of reference and the velocity vector of a static observer at $r$ is 
\be
V^\mu=\frac{u\delta_v^\mu+v\delta_u^\mu}{4ME}.
\ee
The denominator becomes imaginary below the horizon ($r<r_S$) what is in agreement with the
nonexistence of static observers within a black hole. The electromagnetic 
signal $k^\mu$  as received by an observer whose velocity is $V^\mu$ has  a frequency
\be
\omega=V^\mu k_\mu.
\ee

Applying this one finds that the signals emitted (e) from ms and recorded (r) by Alice are redshifted. Their frequency ratio 
turns out to be a decreasing function as the horizon is approached. Its characteristic property is that it may be 
expressed via Alice's speed, $v_A$ as measured by a static and local observer,
\be\label{msaah}
f(ms\to A)=\frac{\omega^r(A)}{\omega^e(ms)}=\frac1{1+v_A}.
\ee
The signals of Alice received by ms are also redshifted,
\be
f(A\to ms)=\frac{\omega^r(ms)}{\omega^e(A)}=1-v_A.
\ee
Both observers can (in principle) infer $v_A$ \cite{Radosz} which approaches the speed of light and 
$f(ms\to A)\to1/2$ as Alice reaches the event horizon. Ms records Alice's signals as 
critically redshifted, 
$f(A\to ms)\to0$ when $v_A\to1$. Hence, from ms's  point of view (as well as from the point of 
view of any other static observer), Alice disappears from the screens as a faint object. 
This is a consequence of the fact that it takes infinite (coordinate) time for arbitrary 
geodesic (also null geodesics) to reach the event horizon. On the other hand it takes finite 
time for Alice (Alice's proper time) to reach the horizon. This asymmetry in the notion 
of time between static and in-falling observers (see also \cite{Grib}) may lead to the question 
about some possible anomaly in communication between falling observers in the vicinity of the
event horizon. 

As a first step in discussion of this problem let us consider Alice and Bob, both 
in radial free fall, exchanging  electromagnetic signals. It turns out that in this case, both Alice and Bob find spectral redshift, that takes a simple form:
\bea\label{abovebaab}
f(B\to A)&=&\frac{\omega^r(A)}{\omega^e(B)}=\frac{1+v_B}{1+v_A},\nn
f(A\to B)&=&\frac{\omega^r(B)}{\omega^e(A)}=\frac{1-v_A}{1-v_B}.
\eea
when expressed  in terms of their speeds. The first equation seems to indicate that function 
$f(B\to A)$ is monotonically decreasing but it is not always realized, see Fig. 1.
The second equation is more singular, both numerator and denominator are vanishing at the
horizon and the detailed shape of the function $f(A\to B)$ is rather involved. 
As is illustrated in Fig. 2 Bob is receiving Alice's signals until the very last 
moment of crossing the horizon. This is manifestation of an effect that he runs into a 
``frozen picture'' yielded by Alice's outgoing signals. Therefore, Alice-Bob communication 
is not broken above the horizon but shows an anomalous character: whatsoever was a difference between 
initial time of release of Bob and Alice, for Bob it looks like he is going to collide with Alice's spacecraft at the horizon. 

\begin{figure}
\includegraphics[width=\columnwidth]{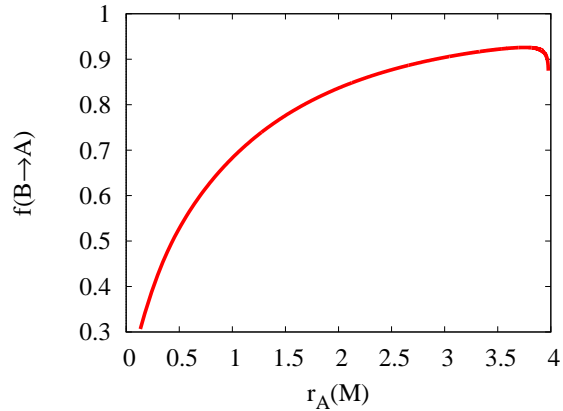}
\caption{Frequency ratio $f\left(B\rightarrow A\right)$ as a function of the radial coordinate of A.}
\end{figure}

\begin{figure}
\includegraphics[width=\columnwidth]{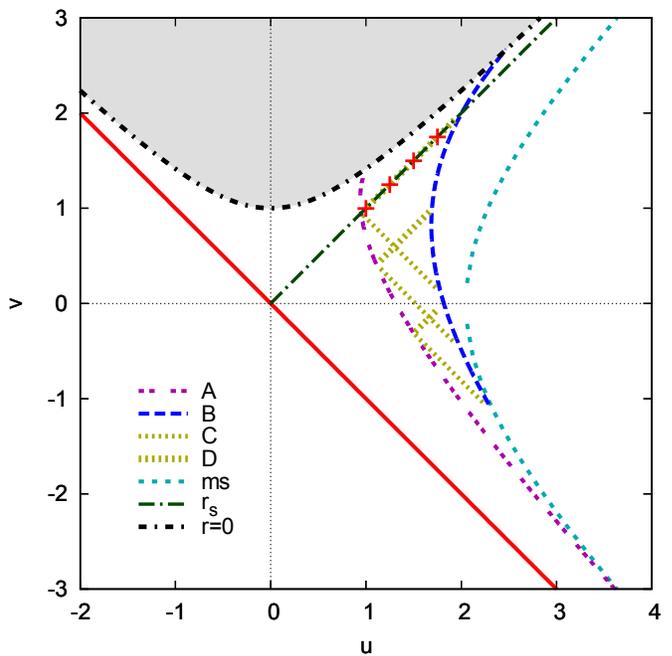}
\caption{Exchange of radial signals above the horizon between A and B. C and D are the in- and outgoing photon geodesics respectively.}\label{above}
\end{figure}

\section{Below the horizon}
Communication below the horizon does not appear to be an obvious extension of the above-horizon communication. In fact, an outgoing radial null geodesic, well defined above the horizon, reveals a singular behaviour at the horizon (see Eq. \eq{out}). That is why neither its extension below the horizon is an obvious issue (the fact that has not been reported yet), nor the term ``outgoing'' could be kept unchanged. Before discussing the problems related to the transfer of the geodesics accross the horizon let us make some general remarks concerning the issue of the arrow of time. 
In the case of a motion of a rigid body crossing the horizon the world-line of a system is oriented and the arrow of 
time \cite{Zeh} identifies the direction of the flow of time in it; the time coordinate 
should be a monotonic function of it. Apart of dynamical context there is another,
geometrical aspect of time. If the space-time geometry is asymptotically flat or if it 
possesses time-like Killing field, the change of the time generates positive invariant length 
square, as opposed to the change of spatial coordinates.

How to find the arrow of time within a black hole, realized by the Schwarzschild geometry? 
This problem on one hand does seem to have a more or less obvious solution: one traces the arrow of time carried by objects falling 
freely into the black hole. On the other hand, one has to pay attention to the fact that global features of test particles
geodesics make the horizon a separatrix, namely there are no geodesics crossing it by traveling away 
from the black hole. One also can verify (see below) that in-going geodesics, massive and massless 
behave in a smooth manner crossing the horizon. These and other arguments, cf. \cite{Boulware} lead to the point of 
view of regular behavior of physical phenomena at the event horizon. Nevertheless, as indicated above (and discussed below) one can observe a peculiarity about the arrow of time on the outgoing geodesics when the two sides of the event horizon are compared.

The discussion here restricted to radial geodesics for the sake of 
simplicity, deals with massive infalling test particle geodesics and two types of massless 
geodesics at each space-time point, one oriented towards the center 
of the black hole and the other directed away, called in- and outgoing null-geodesics, 
respectively as shown in Fig. \ref{above} above the horizon. 

As the horizon is approached, massive and massless ingoing geodesics, given by Eqs. \eq{velkr} and \eq{in},
are continuous functions of the radial coordinate at the horizon (see also a 
special case \cite{Trumper}). It may be shown that also non-radial, ingoing geodesics are regular on the horizon. Light rays are chasing the (infalling) observers 
below the horizon. Ingoing signals, emitted from ms reach Alice beneath the horizon, with a 
redshift
\be
f(ms\to A)=\frac{\omega^r(A)}{\omega^e(ms)}=\frac1{1+\sqrt{1-\frac{h^2_A}{h^2_{ms}}}},
\ee
 This function, valid both in the interior and exterior of the black hole, reproduces the above-horizon formula Eq. \eq{msaah} . At the horizon it takes the characteristic value $1/2$, independent of starting position. As Alice approaches the final singularity, $r=0$, it decreases to 0.
Ingoing signals sent by Bob are recorded by Alice as redshifted,
\be
f(B\to A)=\frac{\omega^r(A)}{\omega^e(B)}=\frac{1+\sqrt{1-\frac{h^2_B}{h^2_{ms}}}}{1+\sqrt{1-\frac{h^2_A}{h^2_{ms}}}},
\ee
This expression is also valid both in the interior and exterior of the black hole, reproducing in this latter case the above-horizon expression \eq{abovebaab}. As $r_A\to0$  it tends to 0.

Above the horizon the outgoing null geodesics, parallel to the horizon, cf. Fig. \ref{above} tend to the line $0\leq u=v$ as $r\to r_S$. However, they display a singular behavior at the horizon $u=v$. Since the arrow of time belongs to geodesics and it is not defined on the separatrix there is no kinematical basis to expect to have it oriented in the same way slightly above and slightly below the horizon. Thus one cannot infer the direction of the time arrow of the  ``outgoing'' null geodesics below the horizon by observations carried out above the horizon. This conclusion remains valid for non-radial outgoing geodesics as well, because the horizon is a separatrix for them, too. The term ``outgoing'' null geodesics will hereafter be applied to geodesics defined as an extension of Eq. \eq{out} for the under-horizon region, $u<v$.

\begin{figure}
\centering
\subfigure[\hskip 3pt Discontinuous scenario. Arrows of time on ``outgoing'' geodesics run antiparallel below and above the horizon. This leads to ``one-way'' A -- B communication: the ingoing signal (dash-dotted line) goes from B to A and the signal from A (dotted line) does not go back to B.]{\includegraphics[width=\columnwidth]{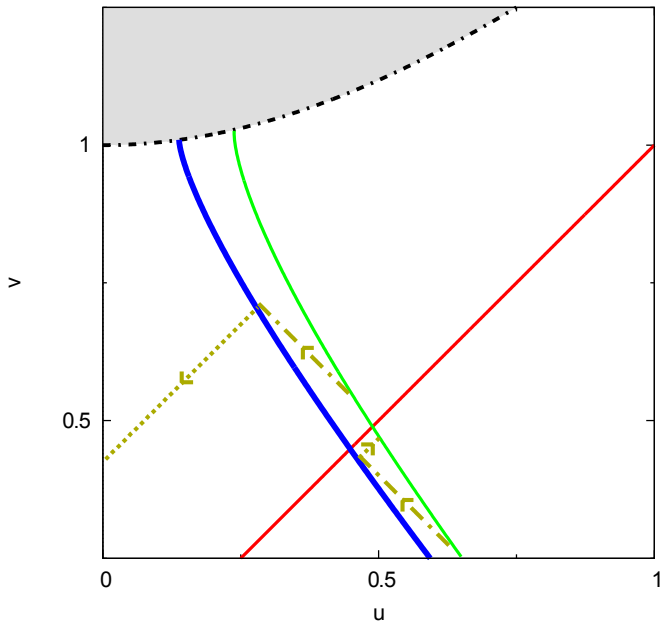}
\label{below1}}\\
\subfigure[\hskip 3pt Continuous scenario. Arrows of time on ``outgoing'' geodesics run parallel below and above the horizon due to discontinuous parameterization. It preserves ``two-way''  A -- B communication: the ingoing signal (dash-dotted line) goes from B to A and A sends a signal back to B (dotted line).]{\includegraphics[width=\columnwidth]{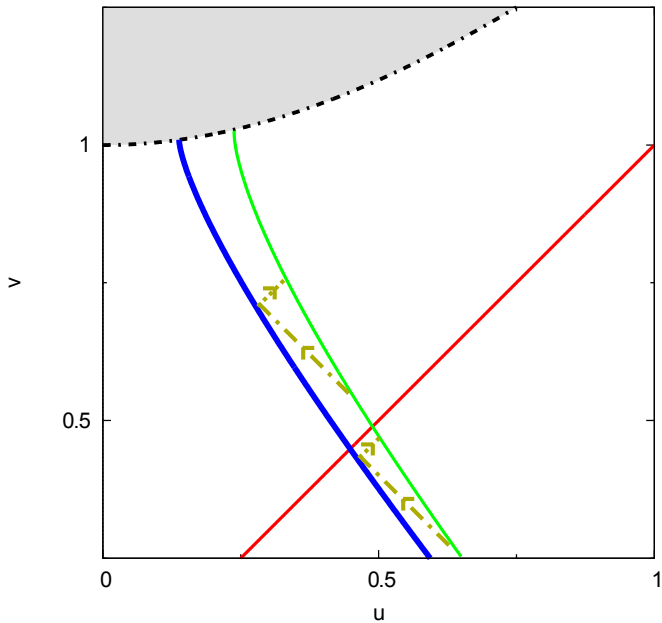}\label{below}}
\caption{Two scenarios of communication below the horizon (the solid line inclined by an angle $\frac{\pi}{4}$) between observers A (thick solid line) and B (thin solid line).}
\end{figure}

The dynamics changes continuously for the regions $r<r_S$ and $r>r_S$ and one may distinguish 
two scenarios related to the fact whether the time dependences in the limit $r\to r_S$ from below and above agree or not.  It is important to realize that the scenarios differ in what the choice of the time coordinate is within the black hole. 

The first scenario occurs for a simple extension of the above-horizon solution for the ``outgoing'' geodesics, 
\be\label{cont}
k_-^v= k_-^u=\frac{4M\Omega}{K(u-v)}.
\ee
for the black hole interior, $u<v$. 
In this case one finds, however, that these ``outgoing'' geodesics \eq{cont}, move away from the center of the black hole (without crossing the horizon) cf. Fig. \ref{below1}. The arrows of time of the outgoing geodesics above and below the horizon run antiparallel and the causal structure suffers a disruptive change (``flips'')  at the horizon. The Schwarzschild coordinate $t$ remains the time as one can check in Fig. \ref{below1}. The geometry within black hole appears to be time-independent. In this case both channels ``in-'' and ``out-'' appear to transfer the signals from B to A - one can refer to this as a ``one-way-communication''. It is the case of continuous parameterization but discontinuous change of the causal structure.

The two-way communication between Alice and Bob may be restored in the black hole interior. But this occurs due to  discontinuous parameterization. Let us suppose that the the parameterization of the ``outgoing'' null geodesics is discontinuous, $\Omega^{below} =-\Omega^{above}$. Then
\be\label{discont}
k_-^v= k_-^u=-\frac{4M\Omega}{K(u-v)}.
\ee
in the range $u<v$. 

``Outgoing'' geodesics as well as ingoing ones approach the center in a monotonic manner cf. Fig. \ref{below}. The time arrows of the outgoing geodesics run parallel above and below the horizon and the causal structure changes continuously at the horizon. Below the horizon the Schwarzschild coordinate $r$ is used then as time, what makes the geometry within black hole time-dependent. 

One can identify a ``signature of time'' as related to continuous or discontinuous change in the causal structure above and below the horizon. In the discontinuous parameterization  the ``signature of time'' does not change i.e. ``time'' related to the arrow of time corresponds to a timelike coordinate, so we will refer to this as a continuous scenario. In the continuous parameterization, the ``signature of time'' flips as ``time'' corresponds to a spacelike coordinate beneath the horizon and we will refer to this case as a discontinuous scenario.

\section{Causal structure of the discontinuous scenario}
The gravitational tidal forces remain regular in either scenario in an extended body 
falling through the Schwarzschild horizon. But the causal structure of the interactions of the elementary
constituents clearly suffer a sudden change sharply at the horizon in the discontinuous
scenario. In fact, if Alice-Bob communication is lost at the event horizon then particles of the extended
system, separated by the horizon cannot maintain their equilibrium position where
action and reaction forces agree. 

To make the loss of the usual causal structure ruling within an extended body more explicit we introduce
the measure of communication $c(a,b)$ for a pair of particles $a$ and $b$, defined by 
1 or 0 when the pair has two- or one-way communication, respectively. The
communication weight 
\be
Q(a)=\frac{\sum_bc(a,b)}{\sum_b1}
\ee
of a constituent $a$ is the fraction of the body which is in two-way communication with 
this particle. A global measure of the causal contact of an extended object 
falling through the horizon is the average of communication weights
\be
\bar{Q}=\frac{\sum_{ab}c(a,b)}{\sum_{ab}1}.
\ee 

\begin{figure}[!h]
\includegraphics[width=\columnwidth]{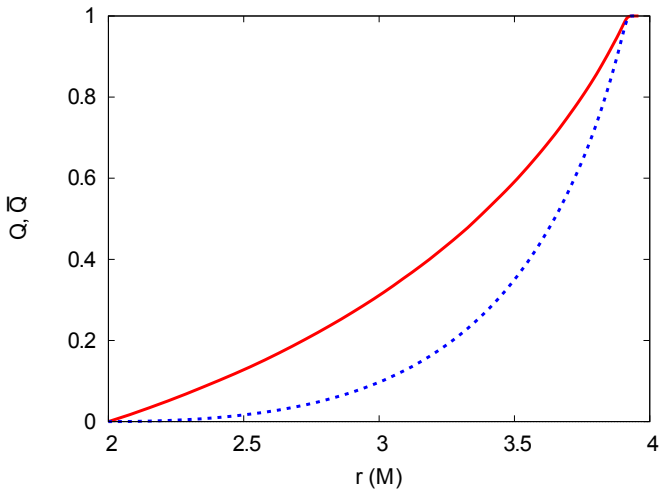}
\caption{Communication weight (solid line) and average communication weight (dotted line) as functions of the radius of the 
outer end of the rod during crossing the horizon.}\label{radius}
\end{figure}

\begin{figure}[!h]
\includegraphics[width=\columnwidth]{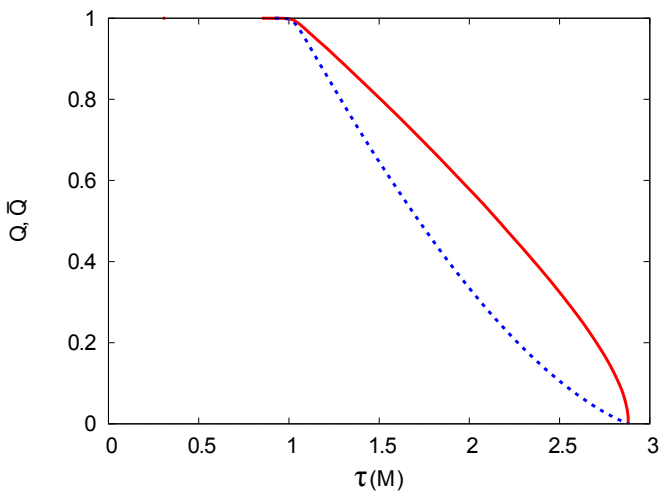}
\caption{Communication weight (solid line) and average communication weight (dotted line) as functions of the proper time of the 
outer end of the rod during crossing the horizon.}\label{prtime}
\end{figure}

Let us discuss here the simplest nontrivial example, a one-dimensional rod, composed of a 
set of weakly interacting point-like particles. The weakness of the interaction allows us 
to approximate the motion of the particles by independent free falls. To simplify the 
discussion further we consider radially arranged rod whose elementary constituents follow 
the geodesics shown in Fig. \ref{below}. The communication and average communication weight, $Q$,
 are depicted in Figs. \ref{radius} and \ref{prtime} as the functions of the radial variable or
proper time, respectively for the discontinuous scenario. Note that any bound or solid 
system of particles requires two-way communication among its constituents, the part of body 
which has already crossed the horizon should fall apart into elementary particles. The figures show
the gradual loss of two-way communication between particle pairs, indicating the gradual 
destruction of extended objects when the horizon is crossed.

\section{Conclusions}
Exchange of electromagnetic signals within Schwarzschild spacetime is discussed in 
the context of communication and interactions in vicinity of the event horizon.
Motivated by the question of a possible scenario of breaking links between two observers 
crossing the horizon in a finite (proper) time, we have studied this problem by means 
of Kruskal-Szekeres coordinates. 

As the communication above the horizon is concerned our main points are the following.
First, observer crossing the event horizon, Alice, can precisely 
identify that critical instant. Alice receives redshifted signals from her mother 
(rest) station. Above the horizon, the frequency ratio is described in terms of a 
universal, i.e. ms position independent function, Eq. \eq{msaah}
which is decreasing with Alice's velocity, tending to $1/2$. At the horizon it
attains its limiting value and in the black hole interior, it drops below 
the horizon's value. Hence, Alice can infer an instant of crossing the event horizon: the 
frequency ratio takes the value $1/2$. Also, it looks like in the black hole 
interior Alice's speed exceeds value c. But `speed' has a sense only above the horizon, 
where it might be measured by a (local) rest observer. Therefore, it is unjustified 
to refer to this case as `the speed greater than c' \cite{Poplawski}. 

Second, exchange of electromagnetic signals between two radially falling observers, 
Alice and (chasing) Bob, may be described above the event horizon in a simple analytic way. 
One finds that both ingoing signals, from Bob to Alice and outgoing signals from 
Alice to Bob are redshifted, being expressed in terms of speeds as in Eqs. \eq{abovebaab}.
In Kruskal-Szekeres coordinates one can observe a special feature of the channel transferring 
outgoing signals above the horizon: Bob receives signals from (preceding) Alice, until 
the very last instant of plunging into the black hole interior. It looks like Bob is 
going to collide with Alice at the event horizon, the actual meaning is that Bob 
recording the signals approaches the spot of their emission. 

More important aspects of communication are related to objects crossing the horizon.
It is found that signals transferred along ``outgoing'' null geodesics, $u=v+C$ display a singularity at the horizon,  $u=v$. This leads to the conclusion that observations made outside the black hole
cannot distinguish between two scenarios. In one of them the signature of time and
causal structure is continuous at the horizon and in the other they change 
in a discontinuous manner. Null geodesics in the out direction show a somehow contrary picture,
they have a natural, continuous frequency parameterization across the horizon in the
discontinuous scenario and a discontinuous realization in the other case.
Continuous scenario corresponds to Unruh's metaphor (see also \cite{Leonhardt}) of a black hole 
interior as a waterfall range bringing both in- and outgoing fish toward the final endpoint.
Quite concrete outcome in this case is the fact that Bob records Alice's signals below their 
releasing spot. But the frequency parameter of the outgoing signals must flip its sign at 
the horizon -- it excludes the possibility of releasing the outgoing signal by observer crossing 
the horizon (such a signal may have been considered as trapped at the horizon). This
contributes to the physical meaning of the presence of the event horizon.

We briefly considered the more exotic, discontinuous scenario in this paper. The two-way communication
is broken in this scenario as the horizon is crossed. The outgoing
null geodesics are parametrized in a natural, continuous manner but signals,
sent by Bob in either in- or out-going directions within the black hole seem to be 
transferred to Alice who is unable to send anything back.

A one dimensional `rod', consisting of an ideal gas of massive particles freely 
falling and crossing the horizon has been briefly considered within the discontinuous scenario.
It was found that a possible interaction among the particles which could establish
a solid ceases to satisfy Newton's third law about the balance of the action and reaction.
In this manner no extended bodies are expected within the horizon of this scenario.
That part of the rod which has already crossed the horizon falls into elementary 
constituents. The horizon appears as a shock wave as it sweeps through the body.

\end{document}